# ARTIFICIAL CONTRADICTION BETWEEN COSMOLOGY AND PARTICLE PHYSICS: THE Λ PROBLEM

Antonio Alfonso-Faus

E.U.I.T. Aeronáutica, Plaza Cardenal Cisneros s/n, 28040 Madrid, SPAIN

E-mail: aalfonsofaus@yahoo.es

**Abstract.** It is shown that the usual choice of units obtained by taking $G = c = \hbar = 1$, giving the Planck´s units of mass, length and time, introduces an artificial contradiction between cosmology and particle physics: the lambda problem that we associate with $\hbar$. We note that the choice of $\hbar = 1$ does not correspond to the scale of quantum physics. For this scale we prove that the correct value is $\hbar \approx 1/10^{122}$, while the choice of $\hbar = 1$ corresponds to the cosmological scale. This is due to the scale factor of $10^{61}$ that converts the Planck scale to the cosmological scale. By choosing the ratio $G/c^3 = $ constant $= 1$, which includes the choice $G = c = 1$, and the momentum conservation $mc = $ constant, we preserve the derivation of the Einstein field equations from the action principle. Then the product $Gm/c^2 = r_g$, the gravitational radius of m, is constant. For a quantum black hole we prove that $\hbar \approx r_g^2 \approx (mc)^2$. We also prove that the product $\Lambda\hbar$ is a general constant of order one, for any scale. The cosmological scale implies $\Lambda \approx \hbar \approx 1$, while the Planck scale gives $\Lambda \approx 1/\hbar \approx 10^{122}$. This explains the Λ problem. We get two scales: the cosmological quantum black hole (QBH), size $\sim 10^{28}$ cm, and the quantum black hole (qbh) that includes the fundamental particles scale, size $\sim 10^{-13}$ cm, as well as the Planck´ scale, size $\sim 10^{-33}$ cm.

Pacs: 98.80.-k, 98.80.Qc

## 1. - Introduction

The appropriate choice of units of mass, length, and time, is a fundamental tool to define the reference scale of the fundamental fields we are dealing with in physics. Reference scales are different from the vacuum state to the cosmological one. Once the units are chosen, pertaining to a particular scale, all physical properties at any other scale are defined as dimensionless numbers: the ratio of the particular property to the defined unit of the corresponding dimension is certainly a dimensionless number. At the cosmological scales the dimensionless numbers should be all of order one. This is something that Dirac proposed in 1937 [1] and we will follow this same line.

## 2. – Planck´s scale

It is well known that the simplest combination of the three constants of physics c (the speed of light, relativity), G (Newton´s constant, gravitation) and ℏ (Planck´s constant, quantum mechanics) defines three units of length, mass and time known as Planck´s units as follows:

$$l_p = (\hbar G/c^3)^{1/2} = 1.6 \times 10^{-33} \text{ cm}$$

$$m_p = (\hbar c/G)^{1/2} = 2.2 \times 10^{-5} \text{ g} \qquad (1)$$

$$t_p = (\hbar G/c^5)^{1/2} = 5.4 \times 10^{-44} \text{ sec}$$

We note that the factor $\hbar^{1/2}$ is present in all these three expressions. It defines the scale. The above units can be derived by equating the gravitational radius of the mass m to its Compton wavelength: $Gm/c^2 = \hbar/mc$. This gives the mass in (1). Then the gravitational radius of this mass is the length in (1), and dividing it by c one gets the time in (1). If we consider the Compton wavelength of m, $\hbar/mc$, as representative of the size of the object, by equating it to its gravitational radius $Gm/c^2$ we have converted the object into *quantum black hole (q.b.h.)*. This is the name given to a particle that has the size, mass and characteristic time of Planck´s units in (1). This particle has an energy $\hbar\upsilon$, similar to that of a photon of frequency $\upsilon$ that has a value approximately equal to the inverse of the Planck time in (1). This energy is also equal to the gravitational potential energy of the q.b.h., $Gm^2/r$. Therefore at this scale we get the junction of electromagnetism, gravity and quantum mechanics. If we now take the units in (1) as the reference when considering any mass, length and time expressed in these units, then these ratios are dimensionless constant numbers that have the limiting value of one as the sizes considered approach the size of the quantum black hole. And then the implication that the units in (1) are the dimensionless number 1 is that $G = c = \hbar = 1$. This is a very widespread choice of units in the scientific literature. We will prove in the next section that the natural value for ℏ at this scale cannot be 1 but $\hbar \approx 1/10^{122}$. The reason is that the reference scale must be the Universe, something that we can contemplate to some extent, and not the inaccessible Planck´s scale.

If we choose the ratio $G/c^3$ = constant and the momentum mc = constant we make constants the factors in front of the action integrals for gravity [2]. The constancy of the factor mc preserves classical mechanics. This ensures to be able to derive the Einstein field equations applying the action principle. The choice of $G = c^3$ allows G and c to be time varying, while the usual choice of G = c = 1 obviously does not allow for any time variation. Then, for any mass m inside a proper volume, the product $Gm/c^2$ is constant. We can state now that: *The gravitational radius $r_g$ of any mass m inside a proper volume of size ~ $r^3$ is a constant.* Taking this radius as the reference to weight the strength of the general relativistic effects in this proper volume we have:

For negligible effects, r >> $r_g$, and for important effects r ≈ $r_g$

We can also say that: T*he Compton wavelength ℏ/mc is constant, because both ℏ and mc are constants, have been taken as constants.*

For a q.b.h. we have in general

$$ℏ \approx mcr_g \approx mc \cdot (Gm/c^2) = (mc)^2 \approx r_g^2 \qquad (2)$$

where we have used the choice $G = c^3$. We see that a q.b.h. has the property r ≈ $r_g$ ≈ $Gm/c^2$ ≈ ≈ ℏ/mc so that the general relativistic effects are important. Also we have for them $Gm^2 \approx ℏc = q^2$. This means that their gravitational energy is of the order of the electromagnetic energy. Here q is the total equivalent electromagnetic charge in the q.b.h. Dividing by the size we get for the q.b.h. energy:

$$mc^2 \approx ℏc/r_g \approx q^2/r_g \qquad (3)$$

Hence, the total relativistic energy of a q.b.h. is of the order of an equivalent photon energy having a wavelength of the order of its size. We know that black holes evaporate, according to the Hawking [3] blackbody radiation emission. For the q.b.h. case we envisage that they do not evaporate: they can be considered as a mass particle, or as a photon, or as a pure electromagnetic energy of the equivalent charge q. It also can be considered as a pure gravitational quantum. It is the junction of general relativity, electromagnetism and quantum mechanics. Considering a Universe closed, unbound and finite (elliptical type) with size L, and mass

M, and since the whole Universe is a proper volume in itself, we have the "black hole" that we call Q.B.H.:

$$GM/c^2 \approx L \qquad (4)$$

and the total momentum content in it is, with $G = c^3$,

$$L \approx Mc = \text{constant} \qquad (5)$$

The constancy of momentum implies the constancy of the size of this Universe, and then we can take the reference

$$L = Mc = 1 \qquad (6)$$

## 3. – Cosmological scale

Calling $\lambda$ the Compton wavelength of a fundamental particle of mass m we have:

$$\hbar \approx mc.\lambda = \text{constant} \qquad (7)$$

Now we may express the mass of the Universe M in terms of a particle of mass m such that $M = Nm$, where N is the total number of such particles. On the other hand we may express the Compton wavelength of these particles $\lambda$ in terms of the size of the Universe $L = n\lambda$, where n is the ratio of scales. Then we may write:

$$Nm.c.n\lambda = M.c.L = (Nn) \, mc.\lambda \approx (Nn) \, \hbar \qquad (8)$$

Since $Mc = L = 1$ and $Nn \approx 10^{122}$ for fundamental particles we get in this limiting process

$$1 \approx 10^{122} \, \hbar \qquad (9)$$

The point is that for the Planck scales it is also $Nn \approx 10^{122}$ so that relation (9) holds true both for the fundamental particle case as well as the Planck´s scale. From (9) it is clear that the value of the quantum Planck´s constant $\hbar$ at these scales must be

$$\hbar \approx 1/10^{122} \qquad (10)$$

and using this value for $\hbar$ in (1) we get

$$10^{61} \, l_p \approx 10^{28} \text{ cm} = L$$

$$10^{61} \, m_p \approx 10^{56} \, g = M \qquad (11)$$

$$10^{61} \, t_p \approx 10^{17} sec = t$$

We see that the scale factor $10^{61}$ is the magic number that converts Planck´s units in (1) into the corresponding cosmological units for the Universe in (11). Then the conversion of Planck´s scale into the cosmological scale is achieved multiplying by the factor $\hbar^{-1/2} \approx 10^{61}$. We note that $10^{122}$ is the number of oscillators of the Planck type required to obtain the thermodynamic properties of the whole Universe as a black hole [4].

Relation (8) implies that for the Universe being $N = n = 1$ the corresponding value of $\hbar$ for this scale is obviously $\hbar \approx 1$. We conclude here that for Planck scale, and the fundamental particles scale, it is $\hbar \approx 1/10^{122}$ while for the Universe $\hbar \approx 1$. While the choice $G = c = 1$ is consistent with our choice $G = c^3$, the choice $\hbar = 1$ is not consistent with the selected scale (Planck´s scale or the fundamental particle´s scale). In our approach the Universe behaves like a self contained QBH and we are inside it.

### 4.- General relation between the cosmological constant $\Lambda$ and $\hbar$

Zel´dovich [5] deduced the following relation for the cosmological constant:

$$\Lambda \approx G^2 m^6 / \hbar^4, \quad i.e. \quad \Lambda \hbar \approx \{(mc)^2 / \hbar\}^3 \approx 1 \qquad (12)$$

Where we have used the relations $G = c^3$ and $\hbar = (mc)^2$. For the cosmological scale we have then

$$\Lambda \approx 1/L^2 = 1, \text{ and } \hbar \approx L^2 = 1 \qquad (13)$$

Fort the quantum and Planck scales we have

$$\Lambda \approx 10^{122} \text{ and } \hbar \approx 1/10^{122} \qquad (14)$$

Then we conjecture the generality of the following relation for any scale

$$\Lambda \hbar \approx 1 \qquad (15)$$

This explains what we call the artificial contradiction of the value of $\Lambda$, the cosmological constant, being of order one as seen from the Einstein cosmological equations, and the value $\Lambda \approx 10^{122}$ as inferred from the standard model of fundamental particle physics. It is just a change of scale.

On the other hand, if $\tau$ is the characteristic time given by the condition $c\tau \approx r_g$, the size of the black hole, $Gm/c^2 \approx c\tau$, we get now the surprising result

$$m \approx \tau \qquad (16)$$

so that mass and time run together. This is the Mass-Boom effect presented elsewhere [2].

## 5.- Entropy

From (2) we see that taking $G = c^3$ we get $\hbar \approx r_g^2$. Hence the $\hbar$ that corresponds to quantum physics is the Planck´s area. It is well known that any black hole of mass M with horizon area $A = (GM/c^2)^2$ has entropy

$$S \approx k\, A/\hbar = k\, GM^2/\hbar c \qquad (17)$$

which is Bekenstein [6] and Hawking [3] formulations for the entropy S of a black hole. In our case with $G = c^3$ and $\hbar \approx (mc)^2$ for the q.b.h. and the Q.B.H. we have

$$S\,(q.b.h.) \approx k_{qbh}$$

$$S\,(Q.B.H.) \approx k_{QBH} \qquad (18)$$

where we have considered the two different values for the Boltzmann constant k. Using the equivalent thermodynamic definition for entropy:

$$S \approx \text{Energy}/T \qquad (19)$$

Energy is proportional to mass and so proportional to the scale factor. For thermal equilibrium, same temperature T everywhere (Planck´s oscillators and the whole Universe), we have

$$S\,(Q.B.H.) \approx k_{QBH} = 10^{61}\, S\,(q.b.h.) \approx 10^{61}\, k_{qbh} \qquad (20)$$

The usual choice of $k_{qbh} = 1$ implies that the entropy of the Universe in these units is just the scale factor presented in (11).

## 6. - Conclusions

The usual choice of units taking the Planck´s constant $\hbar = 1$ is not appropriate for the quantum and Planck´s scales. With the Universe as reference the appropriate value for these scales is $\hbar \approx 1/10^{122}$. The choice $\hbar = 1$ is only valid for the cosmological case. The well known contradiction between the vacuum energy density predicted by the Standard Model of particle physics and the Standard Model of cosmology is explained here as an effect of scales. This is the $\Lambda$ problem.

We conjecture the validity of a general constant product $\Lambda \hbar \approx 1$, i.e., the product of the cosmological constant $\Lambda$ times Planck´s constant, to be of order one for any scale. For the quantum and Planck´s scale it is $\Lambda \approx 10^{122}$. For the cosmological scale, $\Lambda \approx 1$. This explains the contradiction between the values of $\Lambda$ found from fundamental particle physics, the standard model, as compared with the cosmological value. The interpretation for $\Lambda$ is that it gives the number of oscillators (Planck´s quantum black holes) needed to reproduce the thermodynamic properties of the Universe [4], while $\Lambda^{1/2}$ is the scale factor in (11).

## 7. - References.